\documentclass[preprint,12pt,authoryear]{elsarticle}

\usepackage{subfig}
\usepackage{wrapfig}
\usepackage{url}
\usepackage{color,soul}

\usepackage{amssymb}

\journal{ArXiv}

\begin{document}

\begin{frontmatter}



\title{Visualization techniques for the developing chicken heart}


\author[l1,l2,l3]{Ly Phan,
        Sandra Rugonyi,
        Cindy~Grimm
        }

\address[l1]{Intel Corp.}
\address[l2]{Oregon Health and Sciences University, Biomedical Engineering}
\address[l3]{Oregon State University, Mechanical, Industrial, and Manufacturing Engineering}

\begin{abstract}
We present a geometric surface parameterization algorithm and several visualization techniques adapted to the problem of understanding the 4D peristaltic-like motion of the outflow tract (OFT) in an embryonic chick heart. We illustrated the techniques using data from hearts under normal conditions (four embryos), and hearts in which blood flow conditions are altered through OFT banding (four embryos). The overall goal is to create quantitative measures of the temporal heart-shape change both within a single subject and between multiple subjects. These measures will help elucidate how altering hemodynamic conditions changes the shape and motion of the OFT walls, which in turn influence the stresses and strains on the developing heart, causing it to develop differently. We take advantage of the tubular shape and periodic motion of the OFT to produce successively lower dimensional visualizations of the cardiac motion (e.g. curvature, volume, and cross-section) over time, and quantifications of such visualizations.
\end{abstract}


\end{frontmatter}



\section{Introduction} \label{sec:intro}
Cardiac development depends on genetic programs that are modulated by hemodynamic forces and environmental factors. Since the heart starts pumping blood early during embryonic development~\cite{HamburgerHamilton1951,Martinsen2005RefGuide} cardiogenesis mainly occurs under blood flow conditions. Blood flow is essential for proper cardiac development, with altered flow or absence of flow leading to cardiac malformations~\cite{Clark1978,Hogers1999,Hove:2003,Sedmera1999,Tobita2005}. Physically, the interaction of blood flow with cardiac walls generates hemodynamic forces that are exerted on cardiac tissues and that contribute to cardiac tissue deformations during the cardiac cycle. Cardiac cells sense and respond to these hemodynamic stimuli through signaling pathways that alter cell behavior, modulating genetic programs. While at least some of the biological consequences of altered blood flow can be assessed (e.g. resulting cardiac malformation, or changes in tissue composition), and the characteristics of blood flow within the heart measured (e.g. flow velocities, blood pressures), quantifying heart motion and how this motion is affected by altered blood flow conditions poses several challenges. We present here several methods for visualizing and quantifying cardiac motion during early embryonic stages. 

\begin{figure}
\centering
\includegraphics[width=\linewidth]{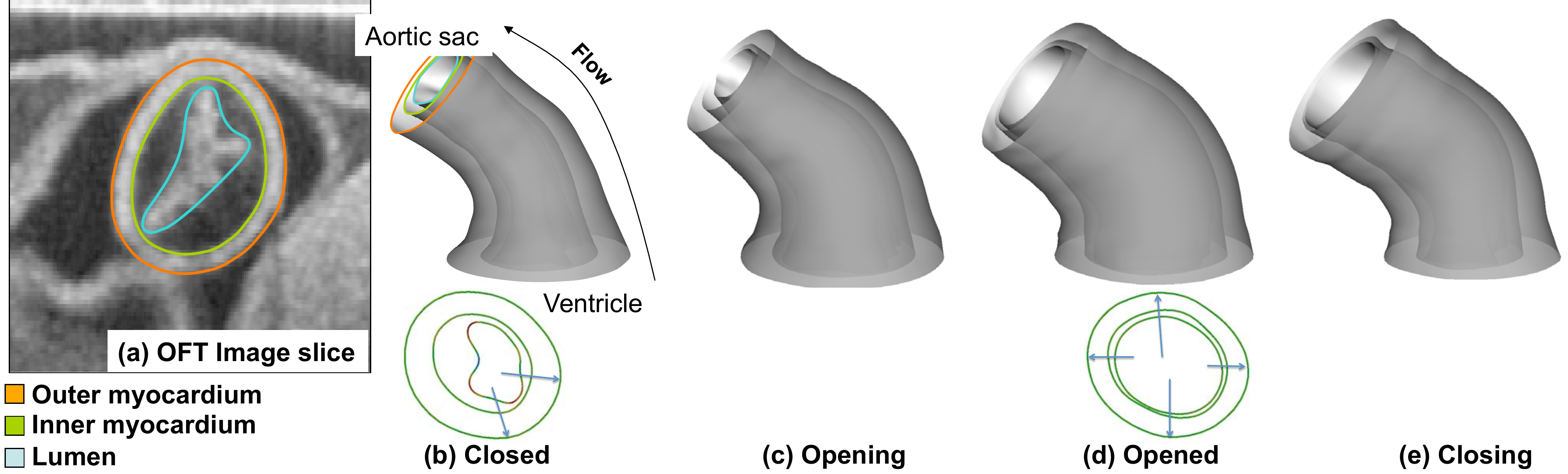}
\vspace{-0.1in}
\caption{
Heart outflow tract (OFT) surfaces reconstructed from OCT data: (a) Original OCT data with contours to illustrate the myocardial walls (external ring-like structure, with the outer and inner myocardium surfaces delineating the myocardium) and lumen-wall interface (bluish line). OFT surfaces depicted when the ventricular end of the OFT is (b) fully closed, (c) opening,  (d) fully opened, and (e) closing.}
\label{fig:OCT-OFT}
\vspace{-0.2in}
\end{figure}

At early stages of embryonic development the heart has a tubular structure. The heart starts as a linear tube that soon begins pumping blood, and then bends and loops to form a looping tubular heart. Chamber and valve formation occur after cardiac looping. At the tubular stages, the heart consists of several segments: a primitive atrium, atrio-ventricular canal, primitive ventricle, and outflow tract (OFT). The OFT connects the heart primitive ventricle to the aortic sac, from which arterial branches emerge. During looping stages heart walls consist of three layers (from outside to inside, see Figure~\ref{fig:OCT-OFT}): a contractile myocardium layer, an extracellular matrix layer (also called the cardiac jelly), and the endocardium monolayer that lines the heart lumen. Developmentally, the OFT is an important cardiac structure because it will later give rise to the interventricular septum, semilunar valves, as well as aortic and pulmonary trunks. Several cardiac malformations therefore originate from the OFT~\cite{Goodwin2014ImpactFlow,Rugonyi2014Malformations,Rothenberg2003BirthDefects}. The early events that lead to these malformations, however, are not well understood. We focus here on the OFT portion of the early developing heart.

\nocite{Goodwin2014ImpactFlow,Rugonyi2014Malformations,Rothenberg2003BirthDefects}

During the looping stages the developing heart and its OFT are very sensitive to hemodynamic conditions. To study the effects of blood flow on cardiac development, embryonic animal models are used, most notably the chicken embryo, which is easy to access, image and manipulate inside the egg. Further, the chick embryonic heart is structurally and functionally similar to the human heart ~\cite{Moorman2003} and has a relatively short developmental time~\cite{Ruijtenbeek2002}. Several interventions to alter blood flow conditions are performed at the looping cardiac stages of chick embryos, one of which is outflow tract banding. In banded embryos, a surgical suture is placed around the OFT and tightened to reduce the lumen area, thus increasing resistance to blood flow through the OFT. Within a couple of hours, banding results in increased blood pressure~\cite{Shi:2013:banding}, increased blood flow velocities through the banded region~\cite{Rugonyi2014Malformations}, and, eventually, cardiac malformations~\cite{Shi:2013:banding,Midgett2014WavePropagation,Rugonyi2014Malformations,Keller1998Embryonic,Yin:submitted}. To analyze cardiac motion under normal and banded conditions, we used 4D images of the chick developing heart OFT acquired using optical coherence tomography (OCT) ~\cite{Liu:2009, AipingLiu2012}. The images were segmented over space and time to extract three surfaces of interest (see Figure~\ref{fig:OCT-OFT}): the outer and inner myocardium surfaces, which line the myocardium layer; and the endocardium or wall-lumen interface ~\cite{Yin:submitted}. We use these surfaces to compare the OFT cardiac motion of normal (n=4) and banded (n=4) embryos.

\nocite{Shi:2013:banding,Midgett2014WavePropagation,Rugonyi2014Malformations,Keller1998Embryonic,Yin:submitted}

%
%
%
%
%
%
%
%

The motion exhibited by the OF is both peristaltic-like (successive cross-sectional contraction) and longitudinal (the tube lengthens and shrinks). The aortic sac end of the OFT is relatively fixed in space, but the ventricular end moves --- and, unfortunately, it moves in and out of the segmented volume. This motion makes consistent parameterization of the tube over time non-trivial. For this reason, we first present an algorithm (Section~\ref{sec:param}) that defines the parameterizations of  the surfaces over time using domain-specific knowledge about how the tissue is deforming {\em and} the desired down-stream analysis. This parameterization also defines how a point on the surface is tracked through time, and allows us to define an approximate correspondence.

Physically altering blood flow by tying a band around the OFT induces cardiac defects, which are thought to depend on mechanical stimuli. Our goal is to quantify normal cardiac motion, and how cardiac motion changes when the band is introduced. Specific questions are: 1) Is the peristaltic-like motion of the OFT walls altered by the band? 2) How does the cross-sectional cardiac shape change during contraction/expansion cycles, and does it change after banding?

Our approach begins by creating consistent parameterizations of OFT surface data sets, first to enable temporal tracking within one chick heart data set, and then to allow temporal alignment of multiple chick heart data sets. Once data sets are consistently parameterized, we apply several data reduction and analysis techniques to examine OFT motion. Specifically, we employ area and volume analysis to produce both qualitative and quantitative measurements of the OFT contraction rate and how contraction travels along the OFT tube, derivative analysis to generate qualitative visualizations of {\em how} the OFT contraction and expansion occur, and curvature analysis to determine how the contraction is influenced by surrounding tissues. 

Contributions: We present a consistent parameterization algorithm suited for analysis of cylindrical-like biological surfaces. Using these parameterizations, we develop several analysis techniques which map complex surface data to 2D images and video streams, and then use standard image processing techniques to generate quantitative plots capturing global behavior.

\section{Related Work} \label{sec:related}
Previous work has been performed in the areas of heart development, parameterization techniques, and visualization approaches for biological data. Here we summarize previous works and their relevance to our study.

\subsection{The developing heart and its outflow tract}
\label{sec:bioStudies}

Early during development the heart is a contracting tube that pumps blood through the embryo circulation. Previous studies have characterized the morphological changes of the heart from a cardiac crescent to the four chambered heart in humans as well as animal models~\cite{Martinsen2005RefGuide,manner:2000,Gittenberger-2005,Moorman:2009}. Further, studies have characterized several aspects of cardiac motion during early embryonic development~\cite{GaritaJHZV2011,AipingLiu2012,Jenkins:2010:Measuring,Keller:1991:Ventricular,Taber:1994:Embryonic}. These studies were mainly descriptive and based on imaged histological sections and in vivo optical imaging protocols. More recently, with the advanced of microscopy techniques and molecular biology techniques, more information is being extracted from cardiac tissues, from spatial localization of cells and molecules, to signaling pathways~\cite{Groenendijk:2005:GeneExp,Runyan:1992:Mediated,Goodwin2014ImpactFlow}. However, analysis of cardiac motion and of subtle changes in cardiac motion are still mainly done by visualization of imaging data.

The OFT is an important portion of the developing heart. In the OFT the cardiac jelly forms endocardial cushions, which are thickenings of the cardiac wall at opposite sides of the tube that generate an elliptical lumen cross-section. This might be an important role of the OFT cushions, since mathematical analyses have shown that an elliptical tube requires less energy for peristaltic pumping than a circular tube ~\cite{TaberPerucchio2000,UshaRao1995}. Studies have also suggested that the elliptical lumen cross-section is facilitated by molecules that tether the endocardium to the myocardium at specific locations, so that as the myocardium contracts, the cushions bulge out away from the tethering spots ~\cite{GaritaJHZV2011}. During myocardial contraction, the cardiac cushions deform until they completely close the OFT lumen by apposition, limiting backflow and thus acting as primitive valves ~\cite{Barry:1948, McQuinn:2007, Patten:1948, Manner:2010}. The motion of the OFT wall during the cardiac cycle and how this motion is affected by altered blood flow conditions, including alterations due to banding procedures, have not been studied in detail and are not well understood. Nevertheless, motion during the cardiac cycle is likely a key component in the mechanotransduction response of the developing heart to hemodynamic conditions.

For the first time, using surfaces extracted from OCT imaging of the OFT over the cardiac cycle, we analyze the motion of the OFT wall under normal (control) and banded conditions. To this end, we used parameterization and visualization approaches. 

\subsection{Parameterization approaches}
\label{sec:related:param}

There are a wide variety of techniques for creating consistent parameterizations of a related collection of surfaces. The simplest approach is to map all the surfaces to a common domain or base mesh~\cite{Hormann2008}. Some parameterization techniques are ad-hoc, but most are based on the minimization of a 2D surface metric, for example, conformal or area measures~\cite{Eck1995,Desbrun2002,Levy2002,DruryVanEssen1997,EssenDDHHA2001,ThompsonEtAl1999,FischlSTD1999}, geodesics on a shape-space manifold ~\cite{CGF:CGF12063,DruryEJM1996}, or strain~\cite{PhanKBRG11}. Alternatively, one can create a deformation of one shape to the other and minimize the 3D deformation energy~\cite{Huang08,ZhangShefferKaick08} or a statistical measure~\cite{Cates06,Datar09}. The techniques that come from the computer graphics or vision literature primarily deal with surfaces that have distinctive features, and aim to map those features on each surface~\cite{Kaick2010}. The OFT surfaces that delineate an expanding and contracting tube-like shape have no such distinctive features, which makes these techniques less applicable. Methodologies that minimize the distortion or deformation from one shape to the next are more applicable. In this work, to achieve a parameterization of the OFT surfaces that will allow proper visualization of the OFT wall motion, we essentially minimize the change in the geodesics of surfaces that represent different phases of the cardiac cycle.  


\subsection{Visualization approaches}
\label{sec:related:viz}

For visualization purposes, the embryonic heart OFT is essentially a deforming tube. Visualizing characteristics of the OFT tube (e.g. curvature or some measure of texture) can be done by color-coding the tube surface and rotating the tube in order to view the tube from all sides. An obvious approach for visualizing the whole tube in a single image is to cut the tube surface open and lay it out flat in the plane, resulting in a 2D image that is easy to inspect and understand. This technique has been used, for example, in virtual colonoscopies ~\cite{ZMKG11,Bartroli:2001:NVC:601671.601736,Hong:2006:CVC:1128888.1128901,Zhao:2008:DEH:1512708.1463032}, in which the primary challenge is to effectively recreate the relatively complex  geometry of the interior of the colon (eg, polyps) while avoiding topology inaccuracies that arise from the scanning process. To better visualize the OFT tube, we employ the same unfolding technique.
Our primary challenge, however, lies in the temporal nature of the data, and thus to achieve proper visualization of the OFT wall motion over time, and of surface-related data (curvature, volume, area) on the 2D images over time. A simple and yet insightful approach is to select several cross-sectional planes along the heart tube and plot either cross-sectional contours or areas within the contours versus time ~\cite{AipingLiu2012,GaritaJHZV2011}. This allows examination of the wall motion along the heart tube and how this motion changes over time and space. However, this approach provides a discrete view of cardiac motion that ignores data in between the chosen cross-sections. We use our approaches to visualize and quantify the motion of the whole heart OFT over space (along the heart tube) and time during the cardiac cycle. Further, quantitative analysis allow comparisons of cardiac motion among normal and banded hearts.

\section{Data acquisition} \label{sec:data}
In this section we briefly describe the procedures used to create the initial 3D surface meshes of the chick embryonic OFT from {\em in-vivo} OCT imaging (fully described in~\cite{Liu:2009,AipingLiu2012,Yin:submitted}). The final output from these procedures, which is used as input for this study, is a set of three 3D surface meshes (outer and inner myocardium surfaces and the lumen surface) at 195 time points over the entire cardiac cycle, which gives a total of $3 \times 195=585$ surface meshes per embryo. At the embryonic stage studied here (approximately 3 days of incubation, HH18), the period of the cardiac cycle is about $400$ ms, and thus the time span between consecutive meshes is about $2$ ms. The set of surface meshes describes the motion of the OFT wall layers (myocardium, cardiac jelly and endocardium - or lumen-wall interface) over the cardiac cycle. 

To prepare embryos for imaging, white leghorn eggs were incubated to the Hamburger-Hamilton (HH) stage 18~\cite{HamburgerHamilton1951}. A portion to the egg shell was removed to gain access to the embryo for imaging. Two groups of embryos were considered: 1) control (n=4), and 2) banded (n=4) embryos. Control embryos were normal embryos in which no intervention was performed. In banded embryos a surgical suture was passed around the OFT, close to the ventricle outlet, and tightened to constrict the OFT wall. Banded embryos were imaged within $2$ hrs of the banding procedure. Both groups of embryos were imaged using OCT at the same developmental stage (HH18).

Our OCT system captured 2D tomographic images at a rate of 140 frames/sec (about 60 frames per cardiac cycle). To get 4D images, 2D cross-sectional OFT images were captured over time for about 2 seconds before moving to an adjacent OFT location. This procedure was repeated until the OCT scanner swept over the entire OFT, and then 2D images of a longitudinal-section were acquired over time (this is needed to fine-tune the reconstruction of the images in the presence of peristaltic motion). These 2D image sequences were then synchronized using image post-processing to reconstruct a 4D image of the heart OFT~\cite{Liu:2009}. Reconstructed images were visually and quantitatively analyzed to ensure proper reconstruction. 

4D image sets were then segmented using a semi-automatic procedure~\cite{Yin:submitted}. Segmentation started by manual tracing of the myocardium layer from a cross-section near the OFT inlet. Using the tracing, 40 tracking deformable elements were generated around the myocardium layer. Each tracking element then adjusted its width and position to fit within the myocardium layer, delineating the myocardium inner and outer contours. The tracking elements were then moved over space (to the next cross-section) and allowed to track the myocardium. The procedure was repeated over space (spanning the entire 3D image) and then over time (spanning the 4D image over the entire cardiac cycle). This procedure generated the myocardial surfaces. The lumen surface was extracted by a similar approach, but the tracking elements were constraint to lie inside of the traced myocardium inner surface. Extracted contours (for each cross-section) were smoothed using a snake-like approach, followed by surface and time smoothing using an average-like procedure. This segmentation procedure provided the 3D surface meshes of the OFT over time used in our study.


In this paper we describe our visualization and analysis techniques and demonstrate them on a small data set (four normal and four banded embryos). Rather than making strong claims about how the data relates to biological function and how it differs in the banded case, we focused on determining whether the quantitative and visualization approaches presented have the potential to differentiate (and quantify the variation in) the data sets.

\section{Consistent parameterization} \label{sec:param}
In this section we describe how we generate a consistent mesh parameterization of the heart surfaces both temporally within one chicken embryonic OFT, and across multiple chicken embryos. We are interested in tracking the motion of the OFT over time. To this end, we take two approaches: 1) we track motion in a space bounded by two fixed end surfaces (OFT inlet and outlet surfaces), thus we use an Eulerian domain (a domain that is fixed in space); and 2) we track tissue particles as they move on the surface (including in the longitudinal tube direction), thus we use a Lagrangian domain (which changes over time moving with the particles). The Eulerian approach is of course simpler to implement given our images and the lack of geometrical features to align the surfaces; the Lagrangian approach is feasible taking into account tissue incompressibility and thus preservation of tissue volume over time. While the Eulerian parameterization allows visualization of wave-like propagation in the tubular heart, the Lagrangian parameterization supports separation of circumferential and longitudinal contraction.

\begin{figure}[t]
\center{
\includegraphics[width=\linewidth]{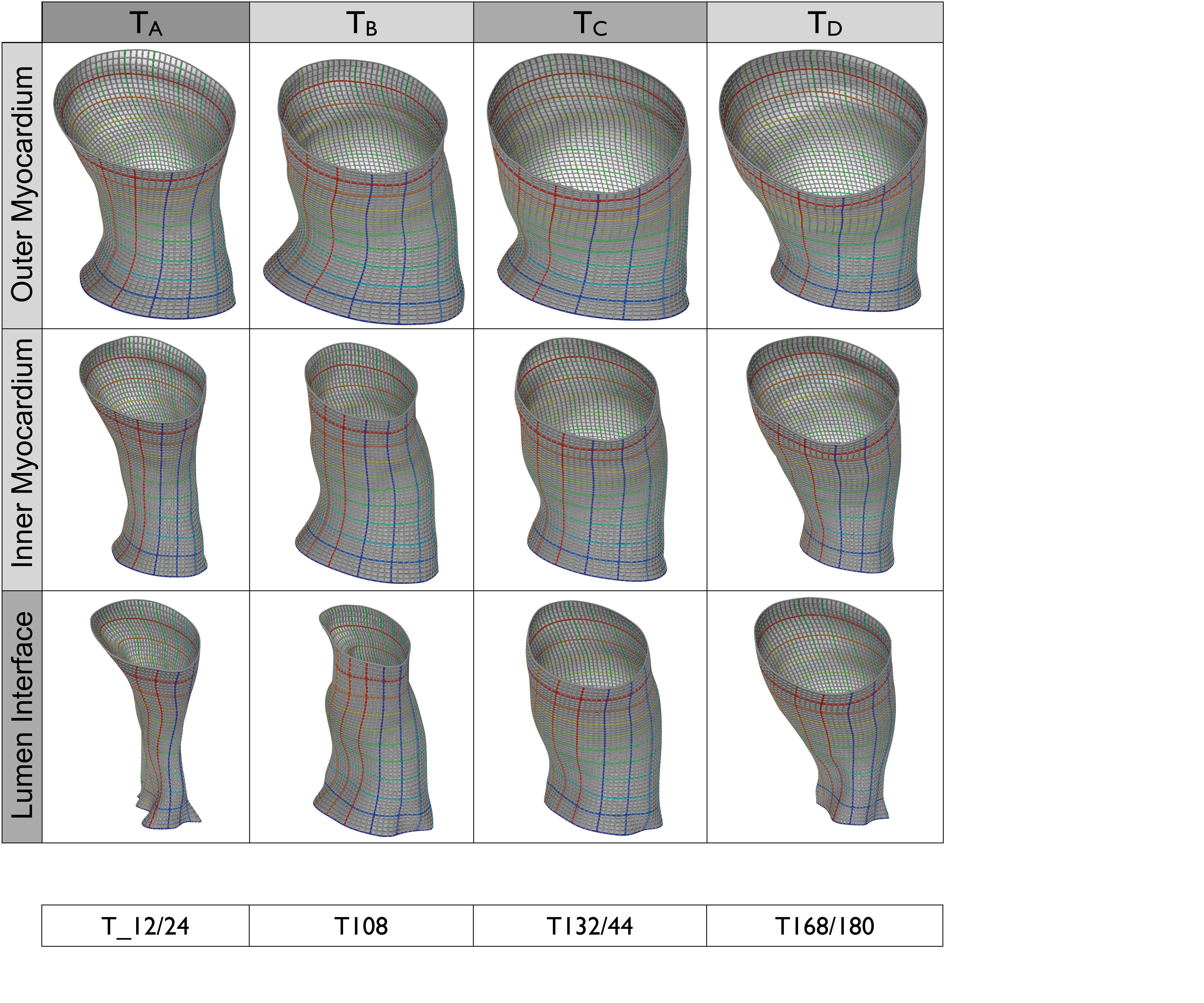}
}
\caption{
Producing a consistent parameterization of the OFT tube at four time points ($T_A $ - minimum, $T_C$ - maximum, $T_B$ and $T_D$ are mid-way between the two). The geodesic constraint results in a relatively uniform geodesic spacing of the grid lines. The inner curve of the tube (darkest blue) is used to align the grid radially between time points and to prevent ``twisting'' of the parameterization. }
\label{fig:param}
\vspace{-0.25in}
\end{figure}

\begin{figure}[t]
\begin{center}
\includegraphics[width=\linewidth]{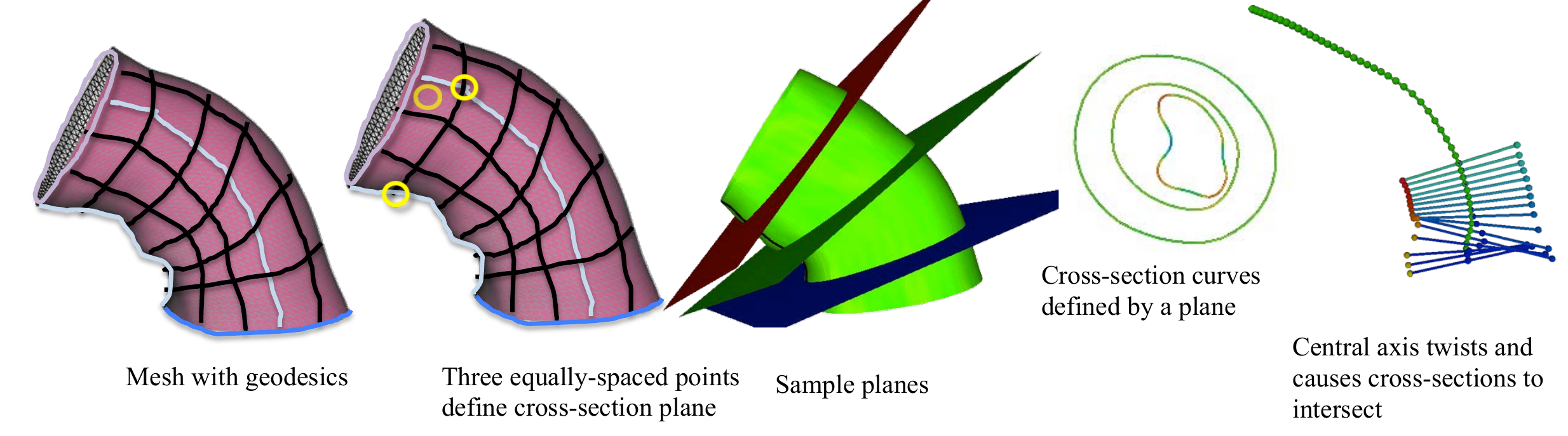}
\caption{Finding consistent cross-sectional planes that do not intersect. a) Sketch of the OFT mesh with the geodesics highlighted. b) Three equally spaced points along a circumferential geodesic, which define a cross-sectional plane. c) Intersection of cross-sectional planes with the surface mesh at three locations along the OFT. d) Cross-sectional contours for the three surfaces in a cross-sectional plane.  e) Using the centerline of the mesh results in planes that self-intersect within the OFT.}
\vspace{-0.25in}
\label{fig:cross}
\end{center}
\end{figure}

Motivations and observations: While during the cardiac cycle the OFT tube moves in space, it remains roughly C-shaped. The longitudinal geodesics (traced from one end of the tube to the other) have a distinct minimum length along the inside of the C-shape (see Figure~\ref{fig:OCT-OFT} for side view). Further, the myocardium can be assumed to be incompressible, which enables estimation of tissue motion given that the aortic sac side of the OFT is fixed in space. Let ${M_t}$ represent the set of surfaces extracted from the 4D images by the segmentation algorithm. Informally, we want A) a geodesic parameterization of these surfaces, B) to use the curve on the inside of the tube (shortest geodesic length) as a stable feature for alignment, C) cross sections along the tube that do not intersect, and D) for the Lagrangian domain, to use the volume of the myocardium to define a consistent parameterization that tracks tissue motion. More formally:

\begin{itemize}
\item[A] Let $u,v \in [0,1] X [0,1]$ be the parameterization of the surface at each time, and let $G_t$ be the mapping from the domain $[0,1] X [0,1]$ to the surface $M_t$ at time $t \in [0,1]$. Specifically, $u$ is the circumferential parameter (around the tube) and $v$ is the longitudinal parameter (along the tube). Then even increments in $u$ and $v$ result in even (geodesic) increments along the surface.

\item[B] For all time $t$, the curve $G_t(0,v)$, $v \in [0,1]$, corresponds to the shortest longitudinal geodesic on the surface $G_t$. By shortest longitudinal geodesic we mean the shortest geodesic from the class of geodesics that connect a point on the inlet boundary to a point on the outlet boundary.

\item[C] Define a cross section at $V(v), v \in [0,1],$ by intersecting $G_t$ with the plane defined by the three points $G_t(0,v)$, $G_t(1/3,v)$, and $G_t(2/3,v)$ (essentially three equally-spaced points on the cross-section contour at $v$). Further we impose that the family of cross sections $V$ for all $v$ do not intersect inside of $G_t$ (see Figure~\ref{fig:cross}). 

\item[D] For the Lagrangian approach, determine how much of the initial surfaces $M_t$ to use to follow tissues (i.e., where to "clip" the OFT end that is close to the ventricle) at each time $t$ assuming that the myocardium volume (encompassed by the outer and inner myocardium surfaces) remains constant. This step is optional, and used here only for the contour analysis (Section~\ref{sec:CSViz}).
\end{itemize}

Recall that, at each time $t$ we have three surface meshes representing the OFT wall: the outer and inner myocardial surfaces, and the lumen surface. Our algorithm creates a single, deforming mesh $G$ for each of those three surfaces such that the {\em shape} of $G_t$ matches the original surface $M_t$ for each time step, but the {\em parameterization} of $G$ has the properties given above. 

We represent $G$ using a mesh that has a grid structure, with $n \times m$ mesh nodes, with $n$ nodes aligned around cross-sectional contours, and $m$ nodes in the longitudinal direction, so that $m$ is also the number of cross-sectional intersecting planes selected for generating the mesh. Please note that cross-sectional planes are not necessarily perpendicular to the heart medial axis. The mesh is created by minimizing differences in geodesic distances among neighboring nodes, while ensuring that cross-sectional planes $V$ do not intersect within $G_t$ (see Figure~\ref{fig:cross}).  In our Eulerian approach the minimization is carried out between the two fixed end planes. In the Lagrangian approach, we adjust the ventricular end of the deforming mesh (allowing it to ``slide'' along the original geometry) so that wall volume is preserved over time. The minimization is then performed after end adjustment. We use the Eulerian approach for all of the visualizations except for the analysis of the contour shapes (Section~\ref{sec:CSViz}).

The minimization leading to the mesh parameterization should be done for each of the three OFT surfaces so that the outer and inner myocardial surfaces, and lumen surface, remain in correspondence. In practice, we parameterize the outer myocardial surface and then project the parameterization to the inner myocardial surface (for each vertex we found the closest point on the inner surface). Because the lumen surface folds, lumen parameterization must be done separately (the projection is not unique). Instead, we parameterize the lumen surface similarly to the outer myocardial surface, {\em except} we align the $v=0$ shortest geodesic (the blue longitudinal line in Figure~\ref{fig:param}) to the shortest geodesic from the outer myocardial surface. Again, the projection of the shortest geodesic curve from the outer myocardial surface to the lumen may not be unique; instead, we project just the end-points onto the two end boundaries of the lumen surface. We then connect these two points with the shortest geodesic.


In what follows, we describe in more detail the algorithms for creating a deforming mesh for a single chick heart OFT over time, and for aligning deforming meshes between two (or more) chick heart OFTs. The parameterization itself provides geometric correspondence.

\subsection{Single OFT alignment algorithm}
\label{sec:singleAlignment}

Our input is a set of 195 surfaces, $M_t$, consisting of outer myocardial meshes that are not consistently parameterized. For each $M_t$, we find the shortest geodesic connecting any two end points on the tube boundary (see (B) above).

We cut the mesh along the shortest geodesic and map it to the unit square using Desbrun's parameterization~\cite{Desbrun2002} with area weighting. We place the cut line so that it aligns with $u=0$, satisfying (B) above. This mapping defines the parameterization $G_t(u,v)$. It approximately satisfies the desired geodesic constraint (A), in that the mapping attempts to preserve area. (C) is satisfied by the construction of the planes from three equally-spaced points on the contour with constant $v$, provided the original surface $M_t$ does not fold back on itself longitudinally in space. Informally, if the curves $G_t(0,v)$, $G_t(1/3,v)$, $G_t(2/3,v)$ do not fold back on themselves then the sequence of triangles defined by those points will not intersect with each other.

To construct our deforming mesh from the parameterization we then place a uniformly-spaced grid (80 nodes in the circumferential direction, 50 in the longitudinal) on the flattened mesh and re-sample the mapped geometry to get the locations for the mesh at time $t$. 

Following this procedure, however, the shortest geodesic can ``slide'' on the surface between time frames because of slight changes or noise in the tube geometry. To address this, we stabilize the {\em end-points} of the $v=0$ cut line from frame to frame, then calculate a new shortest geodesic between those two points. Because the surfaces are fairly similar between frames, stabilizing the end points is sufficient to stabilize the entire $v=0$ cut line. Specifically, to avoid sliding, we take the initial shortest geodesics and temporally smooth the start and end points on the surface boundary by filtering their positions with their neighbors in time (average the positions then project the average back onto the boundary). The mesh locations do not move substantially between time frames so this simple projection onto the corresponding $M_{t-1}$ and $M_{t+1}$ mesh boundaries suffices for establishing correspondence between the points. We then recalculate the shortest geodesic between these slightly modified points.

The re-parameterization of the existing meshes is then used to create a single, deforming surface that roughly maintains equal geodesic spacing and is temporally smooth. We next discuss finding consistent temporal and spatial alignment between multiple chick OFTs.

\subsection{Temporal alignment (multiple hearts)}
\label{sec:temporalAlignment}

We have 195 uniform temporal samples for each chick heart (the $M_t$ meshes) over the cardiac cycle. Our goal is to define a starting point $t=0$ that is at the same time in the cardiac cycle for all chicks. The heart-cycle has distinct points of maximum contraction (minimum cross-sectional surface area) and maximum expansion (maximum surface area). Unfortunately, these points do not occur at the same time along the length of the tube, since the contraction wave travels along the OFT. One approach is to align the meshes based on the total volume of the tube, which essentially ignores the contraction wave. The second approach is to align meshes based on the maximum and minimum areas of a corresponding cross-section. The cross-sectional approach, however, is difficult to achieve in practice. This is because, due to the lack of features in the OFT geometries, finding corresponding cross-sections in not feasible and, due to biological variations, the percentage of  contraction/expansion varies from embryo to embryo. Therefore, we used the volume approach and aligned embryos by volume. 


Specifically, for each chick heart we find the point of maximal contraction by finding the outer myocardial surface with the minimum volume. We set this to be the $t=0$ time point. While not perfect, this alignment is enough for our purposes. If we wish to align both the maximum contraction and the maximum expansion points in the cardiac cycle, then we split the cardiac cycle into two segments and map each temporal segment of the cardiac cycle separately for each chick, essentially enforcing that the maximum expansion occurs at $t=0.5$. We used this approach when comparing the average cross section shape in the expansion versus contraction segments of the cardiac cycle (section~\ref{sec:CSViz}).

\subsection{Between chick spatial alignment}
\label{sec:chickAlignment}

Exact structural alignment between multiple chicks is, in general, not possible. This is because imaging varies from chick to chick. Although every effort has been made to consistently image the same portion of the heart, there are no features to align the images. This is compounded by imaging quality, which degrades as the OFT descends into the embryo. Even though we were careful to measure embryos at the same stage of development (HH18), the embryos do differ in size, which also influences how far into the embryo the OFT descends. 

Within these constraints our parameterization can be used to align the OFTs in the radial direction, and provide roughly similar spacing in the longitudinal direction, even if the starting and ending points are not the same. Specifically, we can compare the overall change along and around the tube but we cannot guarantee that these plots image the exact same portion of the tube.

\subsection{Clipping the ventricular end based on myocardial volume}
\label{sec:volumePreservation}


We have described so far our Eulerian approach, in which cardiac tissue motion moves and deforms within two fixed end planes. In our Lagrangian approach, we use the incompressibility of the myocardium and the fact that the aortic sac is relatively fixed in space (it does not move). This allowed us to ``trim'' or ``clip'' the OFT at the moving ventricular end, so that the cross-sectional plane closer to the ventricle approximately tracks tissue points on the surfaces. To this end, we first calculate the volume between the outer and inner myocardial surfaces for all time steps. Then, we choose the time step with the minimal volume. We next trim the meshes so that myocardial volume is the same (corresponding to the minimal volume found during the cardiac cycle). Specifically, the procedure can be described as follows:
Find the $v_c \in [0,1]$ such that the myocardial volume, when the meshes are sliced by the plane $V(v_c)$, is the same as the minimal volume. Re-parameterize the meshes to the reduced domain. We use this approach to compare contours in Section~\ref{sec:CSViz}.

\section{Cardiac motion visualization and analysis}\label{sec:contract}
In this section we describe several visualization and quantification techniques for analyzing the motion of the heart OFT. OFT wall motion can be described as a peristaltic-like contraction wave that starts at the ventricular end and travels towards the aortic sac. Central to analyzing wall motion along the OFT is the ability to extract cross-sections of the OFT tube. 
We define OFT cross-sections as outlined in Section~\ref{sec:param} --- as the planes $V$ formed from the geodesic sampling in the longitudinal direction. Note that our approach ensures that the planes do not intersect, such that our cross-sectional planes are almost but not perfectly perpendicular to the medial axis.   

Specific questions we are interested in answering are: Does the contraction wave travel at a uniform speed? Do contraction and later expansion happen in a similar fashion along the tube, or do they vary from one end to the other? How does the band affect contraction waves?

\begin{figure}[ht]
\begin{center}
\includegraphics[width=\linewidth]{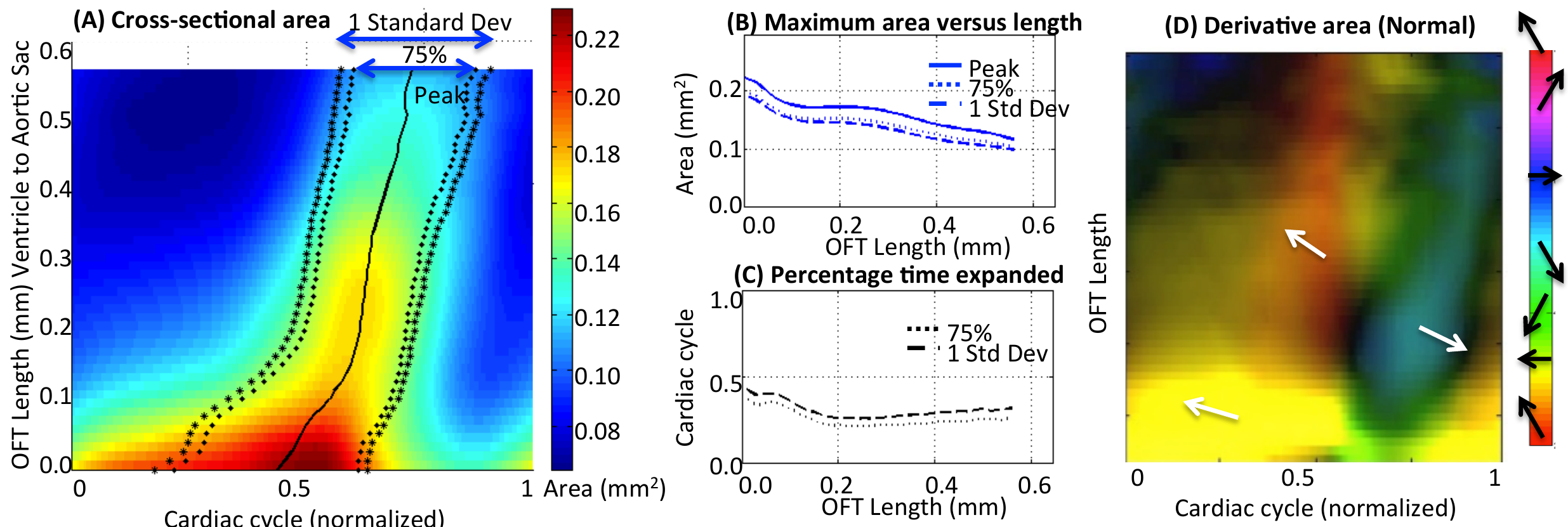}
\caption{Two-dimensional cross-sectional area images and derived quantifications. (A) Cross-sectional area (color coded) as it changes over the cardiac cycle (horizontal direction) and along the OFT (vertical direction). We overlay on the 2D area image the point of maximum area (maximum OFT expansion) along with the points at which the cross-sectional area reaches 75\%  of the maximum (crosses) or one standard deviation from the maximum (stars). (B) 2D plot showing, for each cross-section, maximum area along the OFT. The plot also shows the area that is one standard deviation from the maximum in comparison with 75\% maximal area along the OFT length. (C) 2D plot showing the percentage of time the OFT is expanded, that is OFT cross-sectional area is above 75\% of the maximal area or above one standard deviation from the maximum. (D) The derivative of the cross-sectional area image, with angle mapped to hue and magnitude mapped to intensity.}
\vspace{-0.25in}
\label{fig:areaToLine}
\end{center}
\end{figure}

Using the parameterized meshes, we calculate the internal area of each cross-sectional contour (50 cross-sections at each of the 195 time points, using contours from the three OFT surfaces; see. Figure~\ref{fig:cross}). Because for each surface the cross-sectional areas change both in time ($t$) and space ($v$), areas can be mapped into a 2D image of $v$ (vertical direction) versus $t$ (horizontal direction), mapping areas to colors (blue is smallest, red is largest) and plotting each contour's area as a pixel in the 2D image (see Figure~\ref{fig:areaToLine}A). From these images, we can readily visualize that areas in the ventricle end (bottom) are larger than at the aortic sac end (top), clearly demonstrating that the OFT is a tapered tube. Further, these images enable visualization of the contraction wave that travels through the OFT. 

The 2D area images described above can be analyzed to extract characteristics of the OFT cardiac motion, such as peristaltic wave speed, contraction and expansion rates. Our approach to visualize and quantify cardiac motion is described next.

\subsection{Peristaltic wave visualization} 

We extracted, for each OFT cross-section, the point at which the area reaches a maximum (the middle black line of Figure~\ref{fig:areaToLine}A). These data can be further summarized as 2D plots showing the maximal area along the OFT medial axis (Figure~\ref{fig:areaToLine}B). These plots are useful for comparing OFT sizes and how cross-sectional area changes along the OFT. 


\subsection{Contraction and expansion motion} 

Our next measure examines how long the OFT expansion/contraction lasts, as a percentage of the cardiac cycle. For this measure we tried two approaches, which yielded quantitatively similar results. Our first approach was to use the cross-sectional area data over time (for each position along the OFT), and fit a Gaussian curve  centered on the maximum expansion time for that cross-section. We verified that the data was Gaussian-shaped by visually examining a plot of the data with the fitted Gaussian (see supplemental materials). Then, we found the time at which the area was one standard deviation from the maximum area (see Figure~\ref{fig:areaToLine}A, stars). Our second approach was to use the same data but directly find the time at which the area is 75\% of the maximum  (see Figure~\ref{fig:areaToLine}A, crosses).  These data were plotted together with maximum area along the OFT (Figure~\ref{fig:areaToLine}B). Further, we generated 2D plots showing the time span over which area is expanded, by plotting the percentage of time the area is either above 75\% of maximum or above one standard deviation from the maximum (e.g. Figure~\ref{fig:areaToLine}C). Of course, we can also chose to plot the percentage time in which the OFT area is contracted. These data are useful for comparing whether contraction/expansion timings change from heart to heart and along the OFT length.


\subsection{Contraction and expansion rates} 

As one final visualization we can show the {\em change} in area by taking the image derivative of the 2D area images. The image derivative calculates the gradient of the area, which results in a vector field (vector components are the derivative of area with respect to time, $t$, and the derivative of area with respect to the circumferential direction, $u$). We map the angle of the gradient vector to hue and the magnitude of the vector to intensity  (Figure~\ref{fig:areaToLine}D). From these data, maximum cardiac expansion and contraction rates can also be extracted.

\section{Cardiac shape visualization}\label{sec:shape}
\begin{figure}
\begin{center}
\includegraphics[width=\linewidth]{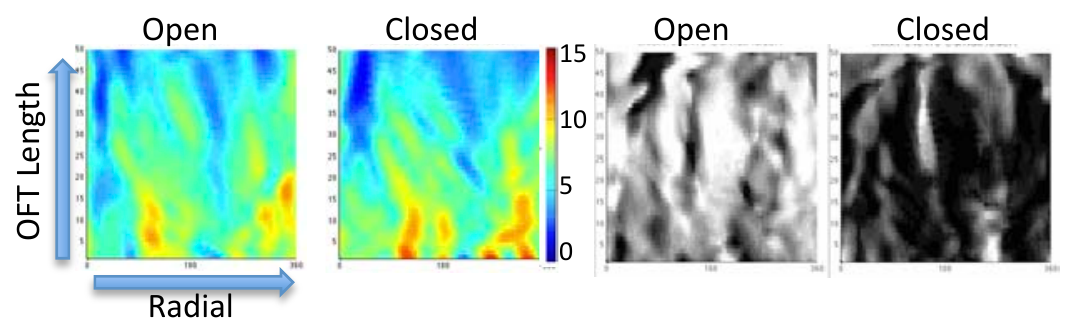}
\caption{2D curvature images. Two frames from a video showing the radial curvature of the OFT outer myocardium at the maximum expansion and minimum contraction points in the cardiac cycle. Right: Normalizing for when the peak curvature occurs (per pixel). See supplemental materials for all videos.}
\label{fig:crossVideo}
\vspace{-0.25in}
\end{center}
\end{figure}

In this section we focus on quantifying the {\em shape} of the OFT and how it changes throughout the cardiac cycle. We first discuss curvature, which is our primary metric for analyzing the shape change along the heart OFT tube, including the lumen. We then describe a visualization approach for the overall change in shape of the myocardium based on strain. Finally, we describe a method that uses a shape space ~\cite{Kurtek:2013:SAM:2500512.2500555} to visualize how the cross-sectional contours deform over the cardiac cycle.

\subsection{Curvature}

We use curvature to elucidate and visualize subtle changes in the heart shape during cardiac motion. Curvature on a surface is a two-dimensional function (the principal curvatures $k_1$ and $k_2$). However, it is common to use either mean curvature (the average of the two) or Gaussian curvature ($k_1 * k_2$) in order to have a single value to display. 

We employed mean curvature for visualizing changes in the lumen surface over time. To this end, we color-coded curvature values and plotted them on the lumen surface mesh. During the cardiac cycle, the lumen surface and its curvature changes, and this change can be visualized on sequential surface meshes or movies showing curvature over time. For the lumen in particular, curvature allow visualization of folds on the endocardial layer.   

When analyzing the shape change of the myocardium, it is necessary to differentiate between curvature along the length of the OFT wall ($v$ direction) and curvature in the circumferential direction ($u$ direction). If we are interested in the contraction/expansion motion of the heart, then circumferential curvature is of more interest. For this case, we explicitly measured radial curvature around cross-sectional contours, ignoring curvature due to the longitudinal bend ~\cite{Xu:2013:CAS:2461555.2461880}.

In either case, surface curvature can be visualized directly on the surface, color-coding for curvature, or can be mapped into 2D images. These 2D curvature images show the color-coded curvature along the tube circumferential direction $u$ (horizontal direction) and along the length of the tube $v$ (vertical direction). This is equivalent to cut the curvature-colored surface open, and map it to the plane using the parameterization (see Figure~\ref{fig:crossVideo}). Changes in curvature over time are then shown using a video or several sequential images. We can further process the curvature images to visualize if all of the points on the surface reach their maximum and minimum curvature at the same time. We can further normalize the curvature of each pixel to the range zero to one using the maximum and minimum curvature for that pixel (see Figure \ref{fig:crossVideo}). Plots of normalized curvature enable to visualize the phase during the cardiac cycle at which maximum or minimum curvature occurs in different regions of the OFT. 

\begin{figure}
\begin{center}
\includegraphics[width=\linewidth]{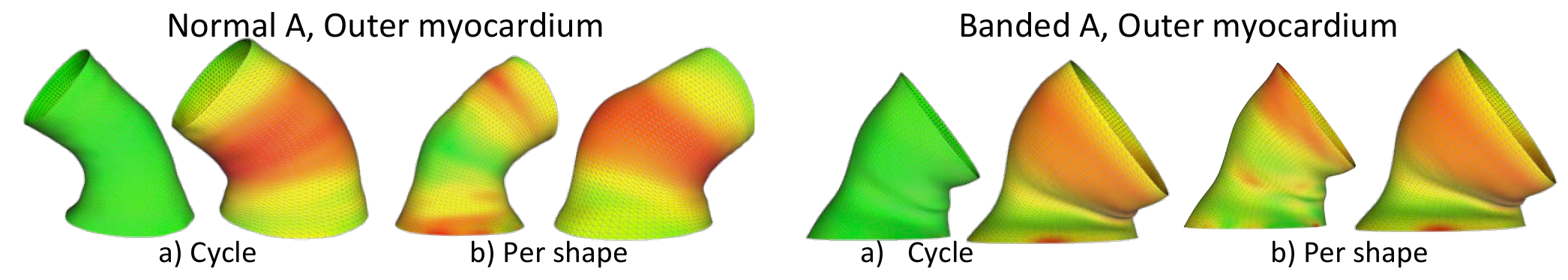}
\caption{ Visualizing the final strain energy $E$ on Normal A and Banded A, outer myocardium. a) and c): The color scheme is normalized across the entire sequence in order to visualize the overall shape change (green-minimum, red-maximum). b) and d): the color scheme is normalized for each time point in order to more clearly show the peristaltic motion. Left image is maximum contraction (closed), right image is maximum expansion (open). }
\vspace{-0.25in}
\label{fig:Strain}
\end{center}
\end{figure}

\subsection{Strain visualization} 
\label{sec:StrainViz}

The OFT heart wall contracts and expands. We can therefore use strain, a measure of surface deformation, to visualize cardiac motion. Strain here is defined with respect to a reference cardiac surface mesh, which we chose to be the surface at maximum contraction (minimum volume). Strain is essentially how much an infinitesimal circular disk in the reference surface expands (or contracts) along two orthogonal (principal) axes ~\cite{PhanKBRG11,Knutsen2010}. To represent (and visualize) strain as a single measure, we sum the square of the normalized changes in length - the strain energy E - and then color-code the surface for strain (see Figure~\ref{fig:Strain}). Strain images of the myocardium surfaces enable better visualization of expansion and contraction waves traveling through the heart OFT.

\subsection{Cross-sectional motion}
\label{sec:CSViz}

\begin{wrapfigure}{l}{0.425\linewidth}
\vspace{-0.5in}
\center{
\includegraphics[width=.8\linewidth]{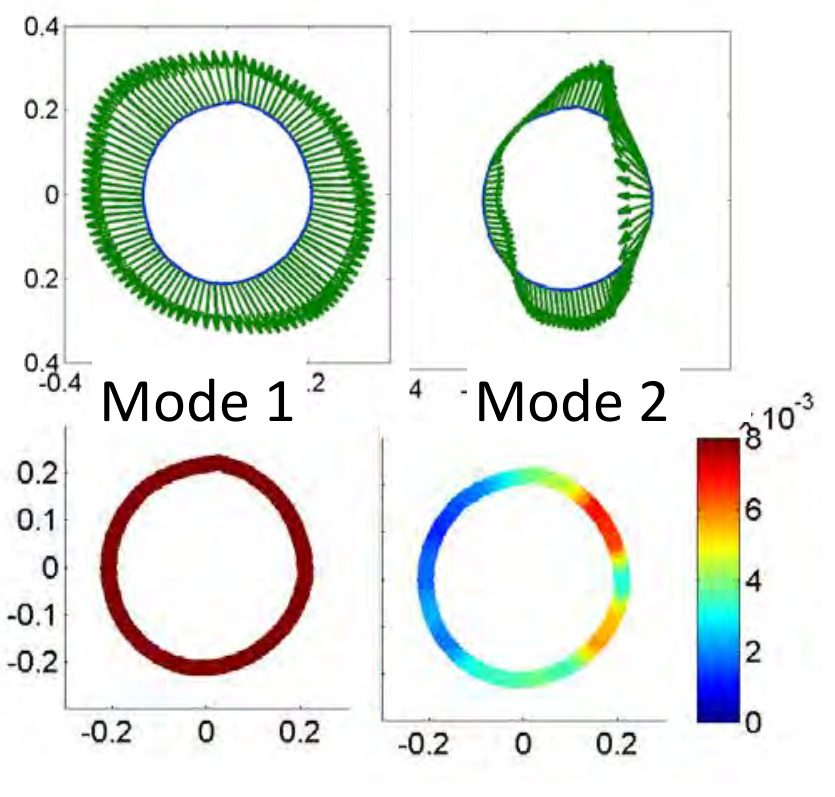}
}
\vspace{-0.1in}
\caption{
PCA of a single contour (contraction phase, ventricle end, Normal A). Left: First mode (size change). Right: Second mode (movement after size change accounted for). Top row: Displacement direction. Bottom row: Displacement magnitude.}
\vspace{-0.2in}
\label{fig:PCAExample}
\end{wrapfigure}

In this section we describe a method for visualizing the shape of the cross sections as they change over time and along the length of the OFT. This technique is the equivalent to performing Principal Components Analysis (PCA) on the space of shape variation~\cite{Kurtek:2013:SAM:2500512.2500555}, analyzing the shape change of individual cross sections~\cite{Kurtek:2013:SAM:2500512.2500555,KurtekCurve}. For this visualization we used our Lagrangian mesh corresponding approach, and we clipped the ventricular end of the meshes so that myocardial volume is preserved. We then separated expansion versus contraction phases for each OFT. We used the approach of Kurtek et al~\cite{Kurtek:2013:SAM:2500512.2500555}, which performs the statistical equivalent of PCA on the set of curves, producing orthogonal modes of variation, which can then be plot as displacement plots from a reference (average) contour (see Figure~\ref{fig:PCAExample}).  Note that the analysis is not temporal so the arrows point in the same direction for both expansion and contraction. These analyses allow visualization of individual cross-sectional shape changes during expansion and contraction of the heart OFT.


\section{Results and discussion} \label{sec:results}




In this section we show results from normal and banded embryonic OFTs, and briefly discuss implications.

\subsection{OFT volume analysis}

A few observations about the heart OFT surface data, including limitations. First, while the OFT tube contracts both circumferentially and longitudinally, our surface segmentations were constrained by two end planes, the inlet and outlet planes. OFT tissues therefore slide in and out of the segmentation window. Second, OCT image resolution and quality decreases with imaging depth. Because the OFT is a curved tube that descends into the egg, the exact end (outlet) of the OFT, marked by the aortic sac, cannot usually be located within the OCT images. Further, the OFT inlet moves substantially with the motion of the ventricle. Thus, tracking of OFT "particles" is not feasible, and can only be approximated. Third, OFT surface segmentation is, on average, within 1-2 pixels of the actual surface (as verified by an expert). This corresponds to about 10 microns, which is approximately 5\% of the maximum OFT radius (when fully expanded). Fourth, the lumen surface is intricate, with tethering points and irregularities especially upon OFT contraction, in comparison with the tubular-like, smoothed myocardial surfaces. Thus myocardium surfaces are  generally more precisely segmented than lumen surfaces. Nevertheless, all surfaces could be used for our analyses. 

\begin{figure}
\center{
\includegraphics[width=\linewidth]{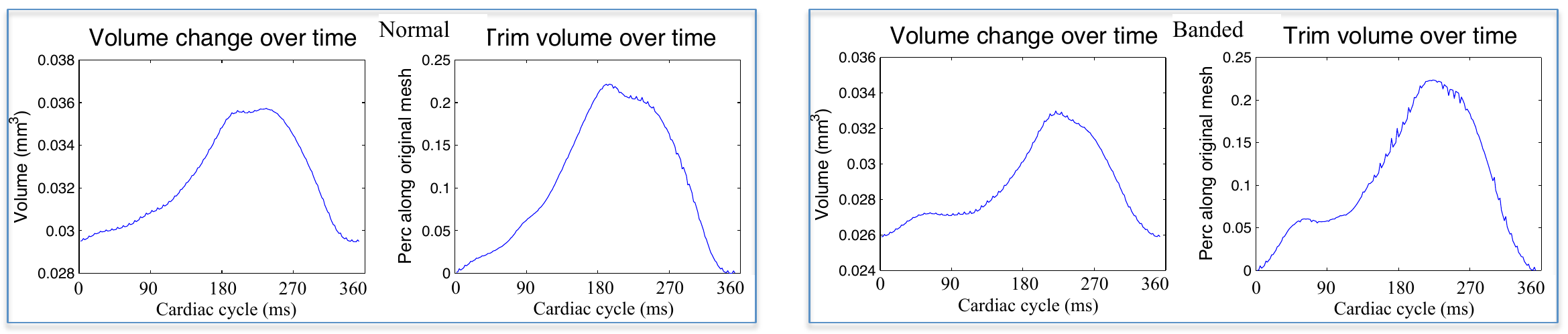}
}
\caption{
Volume normalization for a normal (left) and banded (right) chick heart OFT. Left: The volume between the outer and the inner myocardium in the original meshes. Right: The trim distance along the original mesh (as a percentage of the original length) that results in a constant myocardial volume throughout the cardiac cycle.}
\vspace{-0.2in}
\label{fig:volume}
\end{figure}

Within the uncertainties of our imaging, reconstruction, and segmentation strategies, however, we partially addressed the tracking of the OFT tissues by invoking tissue incompressibility (preservation of volume over the cardiac cycle) for the myocardium layer. By calculating the volume of the myocardium layer (the volume enclosed by the inner and outer surfaces) over time, we can estimate how much tissue slides out of our fixed window. The myocardial volume change over the cardiac cycle is about 20\%, indicating that longitudinal contraction of the OFT tube is about 20\% (see Figure~\ref{fig:volume}). This is in line with our previous observations~\cite{Liu:2009}. Nevertheless, visualization can also be accomplished for a window fixed in space. We have addressed both strategies here. 

Contraction and expansion times differ for each chick heart. When OFT wall contraction and expansion is measured using the volume of the outer myocardium, with expansion representing an increase in volume and contraction a decrease in volume, the percentage of time in the cardiac cycle during which the OFT expands and contracts varies from embryo to embryo (see Table~\ref{tab:volume}). Using the maximum contraction point (minimum volume) as a reference for alignment $(t=0)$, maximum expansion (maximum volume) occurs between $34\%$ and $49\%$ of the entire cardiac cycle. These differences are likely due to biological variations and perhaps small changes in temperature control during image acquisition (temperature is known to regulate cardiac cycle in early chicken embryos). Nevertheless, they reveal that complete alignment of the embryos over time is challenging. 

\begin{table}[h]
\begin{center}
\begin{tabular}{|l|c|c|c||c|c|c||c|c|c|c|c|c|c|}
\hline 
& \multicolumn{3}{|c|}{Ratios} & \multicolumn{3}{|c}{Periods (ms)} & \multicolumn{5}{|c|}{Speeds (mm/s)} \\
\hline
Chick & Expand & Contract & Ratio & Period & Expand  & Contract & min & max & avg & std & cycle \\ 
\hline 
Normal A & 38.27\% & 61.73\% & .61 & 372  & 142 & 230 & 3.3601 & 18.499 & 8.9619 & 3.4245 & 6.8149 \\
Normal B & 34.18\% & 65.82\% & .51 & 347 & 119 & 228 & 3.6898 & 50.872 & 15.485 & 11.095 & 6.2511\\
Normal C & 49.49\% & 50.51\% & .97 & 341 & 169 & 172  & 1.9371 & 62.039 & 13.443 & 9.8212 & 6.6984\\
Normal D & 35.20\% & 64.80\% & .54 & 337 & 119 & 218 & 4.8868 & 40.316 & 13.918 & 6.8177 & 8.5891\\
Banded A & 36.22\% & 63.78\% & .56 & 355 & 129 & 226 & 5.0499 & 47.694 & 17.978 & 10.432 & 9.2953 \\
Banded B & 44.39\% & 55.61\% & .79 & 355 & 158 & 197 & 2.9277 & 12.388 & 7.3956 & 2.3552 & 5.8999 \\
Banded C & 35.71\% & 64.29\% & .55 & 359 & 128 & 231 & 5.3932 & 15.672 & 9.2273 & 2.7269 & 8.0928 \\
Banded D & 49.49\% & 50.51\% & .97 & 350 & 173 &  177  & -143.18 & 2923.4 & 90.975 & 425.65 & 25.259\\
\hline
\end{tabular}
\end{center}
\caption{Left columns: Expansion and contraction ratios for the eight data sets, as a percentage of the cardiac cycle. We show the ratio of expansion to contraction in the fourth column; for some chick hearts expansion takes the same amount of time as contraction, for others, contraction takes longer. Middle: Actual length of cardiac cycle. Right: Speed of the peristaltic wave along the OFT. Minimum, maximum, average and standard deviation of the expansion wave speed along the OFT. Cycle: measure of the speed of the wave considering only the OFT ends.}
\label{tab:volume}
\end{table}%


\subsection{OFT wall motion visualization and analysis}

\begin{figure}[ht]
\begin{center}
\includegraphics[width=\linewidth]{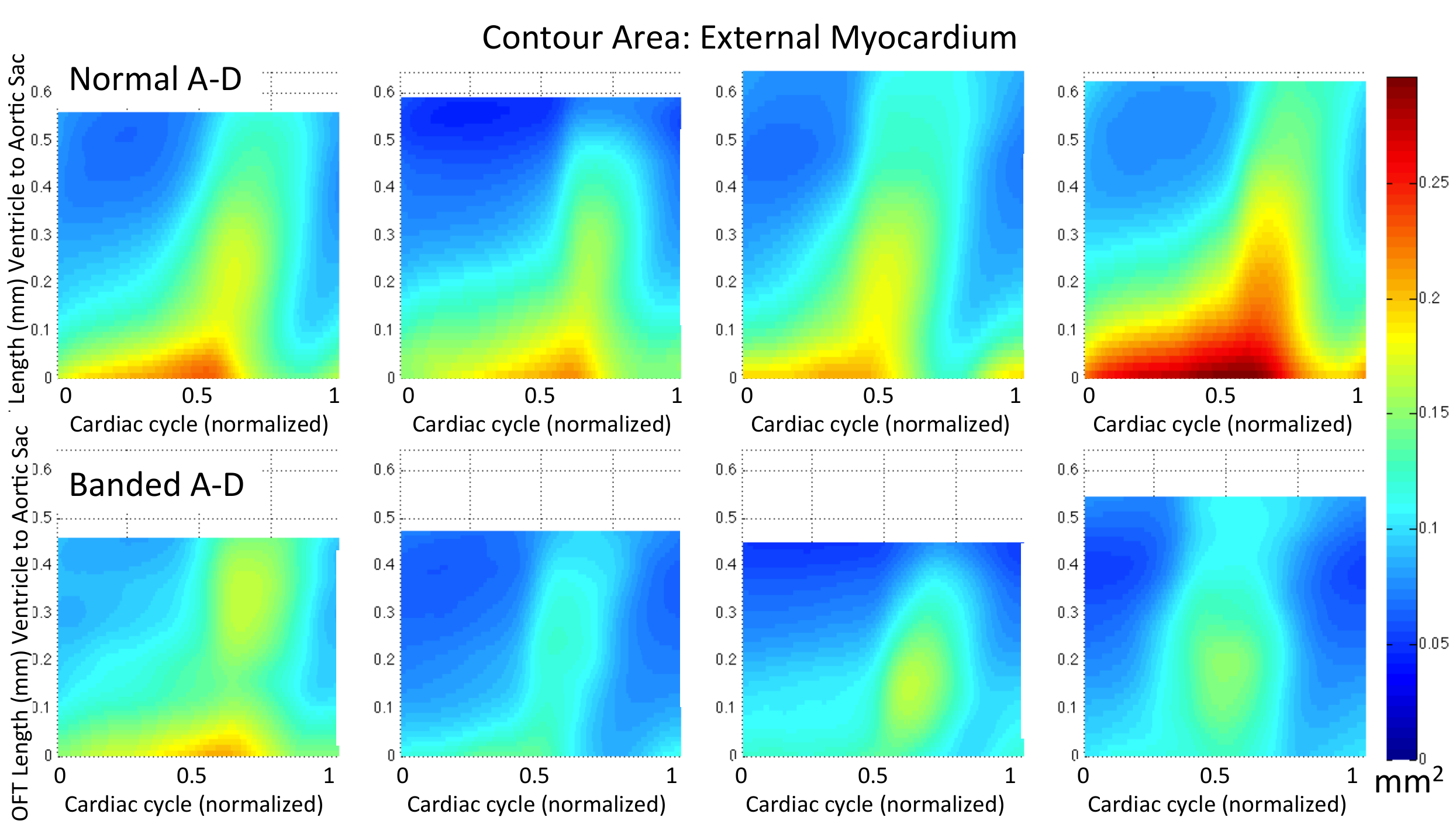}
\caption{2D area images of the OFT. Cross-sectional area of the external myocardial surface  as a function of cardiac cycle ($x$-axis, normalized) and distance along the OFT ($y$-axis in mm). From these 2D area images, OFT wall motion can be clearly visualized. A contraction/expansion wave can be seen as it travels along the OFT. The top four images are from normal embryos, the bottom four images are from banded chicks.} 
\vspace{-0.2in}
\label{fig:area}
\end{center}
\end{figure}

2D area images, which plot cross-sectional areas along the OFT and over time, enable easy comparisons of OFT wall motion in control and banded embryos (see Figure~\ref{fig:area}). The effect of the band is clearly observed from these images --- banded embryos show a ``neck'' region corresponding to the location where the band restricts expansion. Control embryos, on the other hand, show a more gradual change in cross-sectional area along the OFT. Similar results were found for all cross-sectional areas (from the external and internal myocardial surfaces and the lumen surface). The 2D area images, therefore, nicely summarize the effect of the band on OFT wall motion as well as the band position relative to the segmented surfaces.   

\begin{figure}[ht]
\begin{center}
\includegraphics[width=\linewidth]{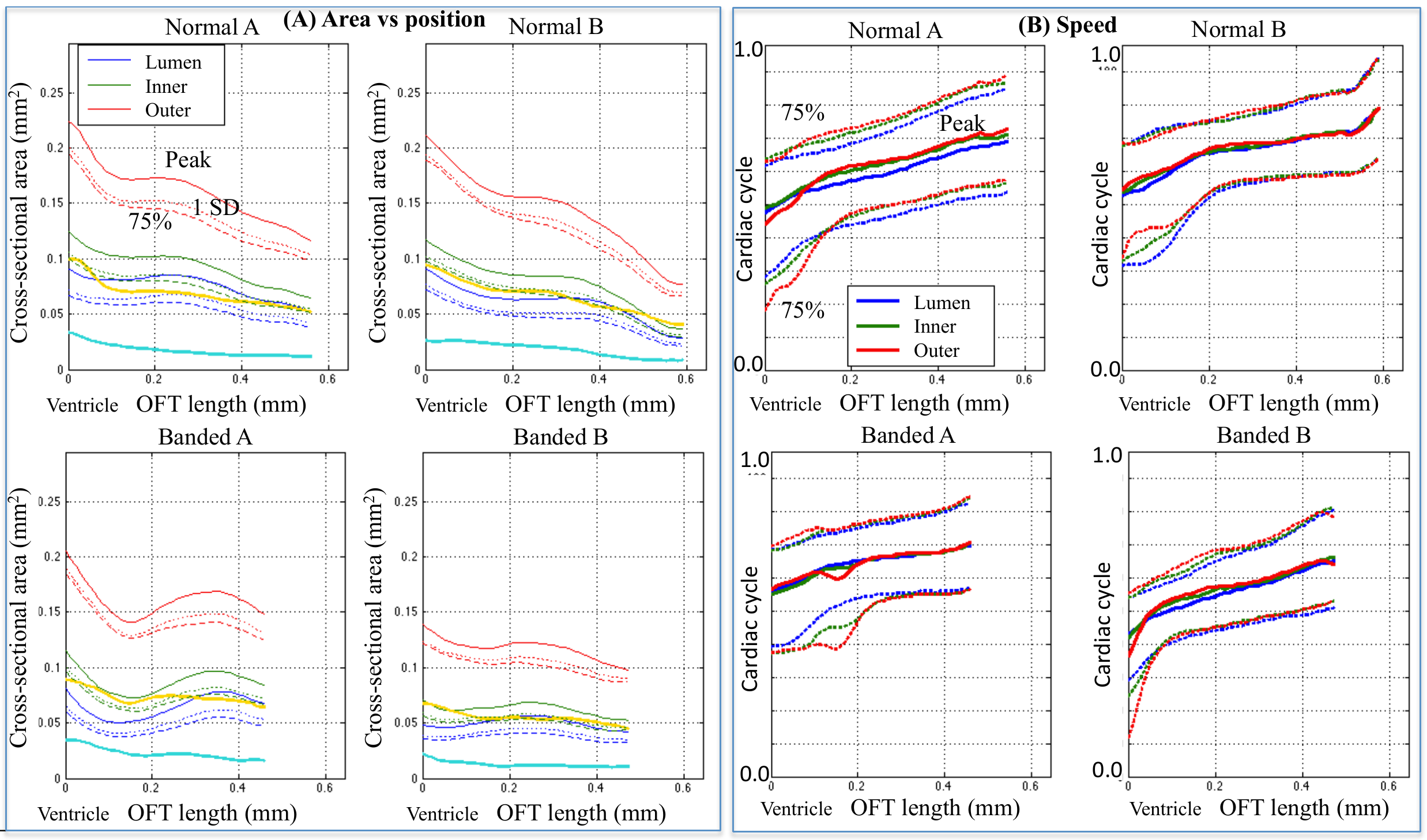}

\caption{Plots of maximum cross-sectional area data for selected normal and  banded chicken embryo OFTs. Cross-sectional area data for the external and internal myocardial surfaces as well as the lumen surface are shown. (A) Left: Cross-sectional area versus OFT length. The plots show maximum area (peak) as well as 75\% of the maximal area and the area corresponding to one standard deviation from the maximum.  (B) Middle: Time at which maximum expansion occurs versus OFT length. The normals have a steep slope followed by a shallower one. The banded examples have a shorter steep slope section and the slopes of the second section are all shallower. }
\vspace{-0.2in}
\label{fig:areaPlots}
\end{center}
\end{figure}

By plotting the maximum cross-sectional area versus position along the OFT (Figure~\ref{fig:areaPlots}A) we can clearly see the effect of the band. In normal embryos, maximum cross-sectional area decreases from the ventricle to the aortic sac end of the OFT, with a plateau around 0.2 mm. In banded embryos, however, maximum cross-sectional area presents a local minimum that depends on the location of the band. Data from the lumen surface as well as the external and internal myocardial surfaces track each other fairly closely in the last portion of the OFT (after the plateau for normal and after the local minimum for banded) but not close to the OFT inlet. 

We can also plot the {\em time} at which the cross-sectional area is maximum along the OFT (Figure~\ref{fig:areaPlots}A). This plot can be used to measure the speed of the cardiac expansion wave along the OFT. For normal chicks the wave does not travel at a uniform speed along the OFT (Figure~\ref{fig:areaPlots}B). The banded chicks have a less distinct pattern, and the slopes of the linear section are all either much shallower or much steeper than the normal chicks, meaning that the wave takes less (or more) time to travel down the OFT . We measured expansion wave speeds for the entire tube (see Table~\ref{tab:volume}). Note that in the banded D case a substantial portion of the tube contracts simultaneously (rather than through a peristaltic-like motion), and, in fact, the wave actually travels {\em backwards} for the initial part of the contraction cycle.

\begin{figure}[ht]
\begin{center}
\includegraphics[width=0.49\linewidth]{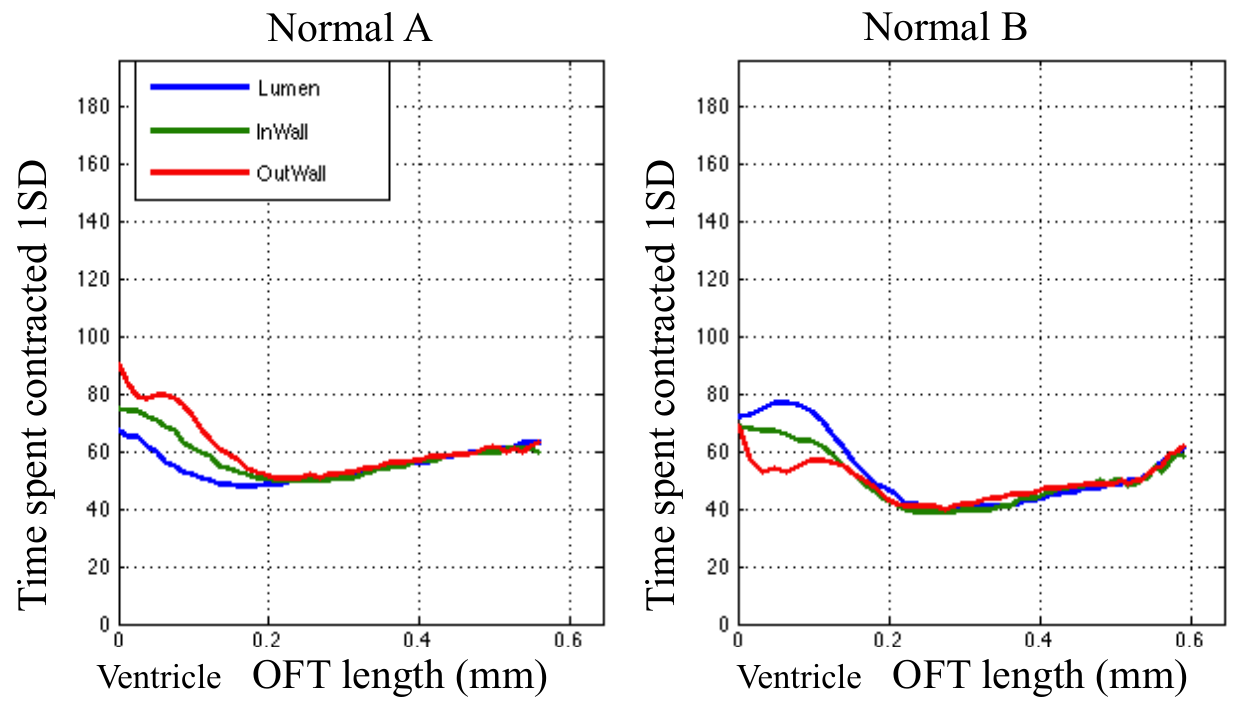} \hspace{0.1in} \includegraphics[width=0.49\linewidth]{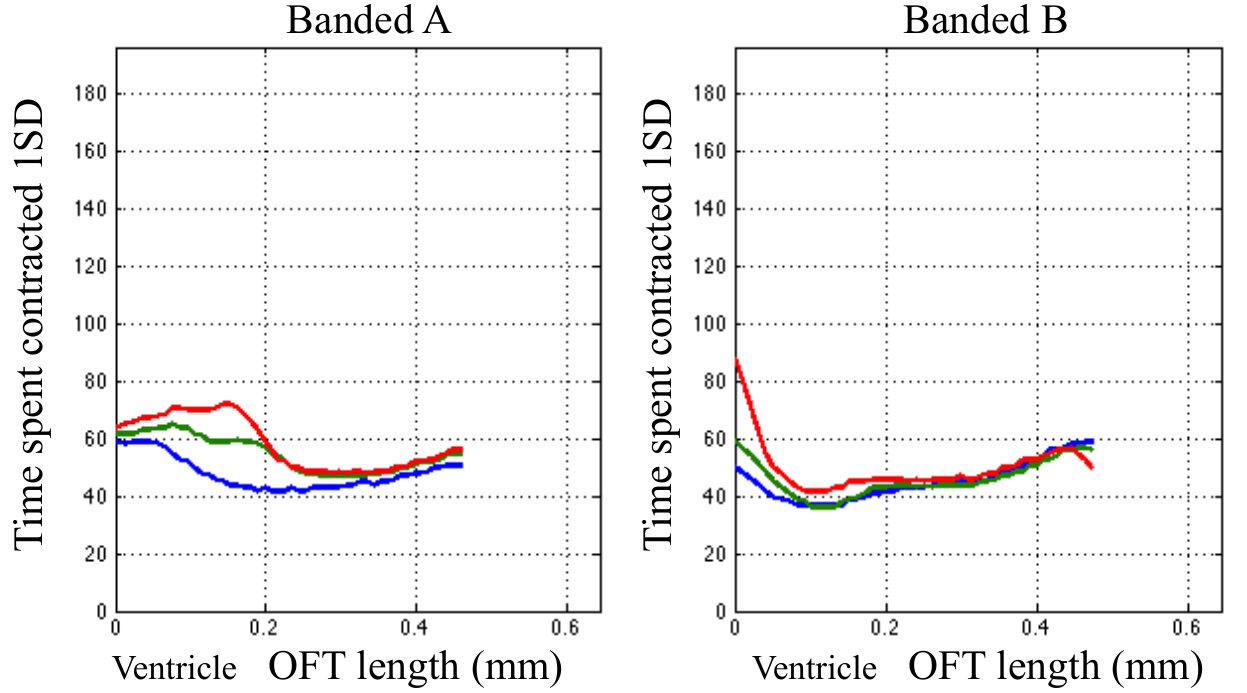}

\caption{Plots of the percentage time expanded (measured as within one standard deviation in area from maximum expansion). Normals are on the left, banded on the right.}
\vspace{-0.2in}
\label{fig:areaPlotsII}
\end{center}
\end{figure}

Further, we compared the percentage time at which cross-sectional areas along the OFT are expanded (Figure~\ref{fig:areaPlotsII}), as measured by being within one standard deviation of the maximum expansion area. The percentage time expanded decreases for approximately the first third of the tube, then steadily increases towards the aortic sac end. 

\begin{figure}[ht]
\begin{center}
\includegraphics[width=\linewidth]{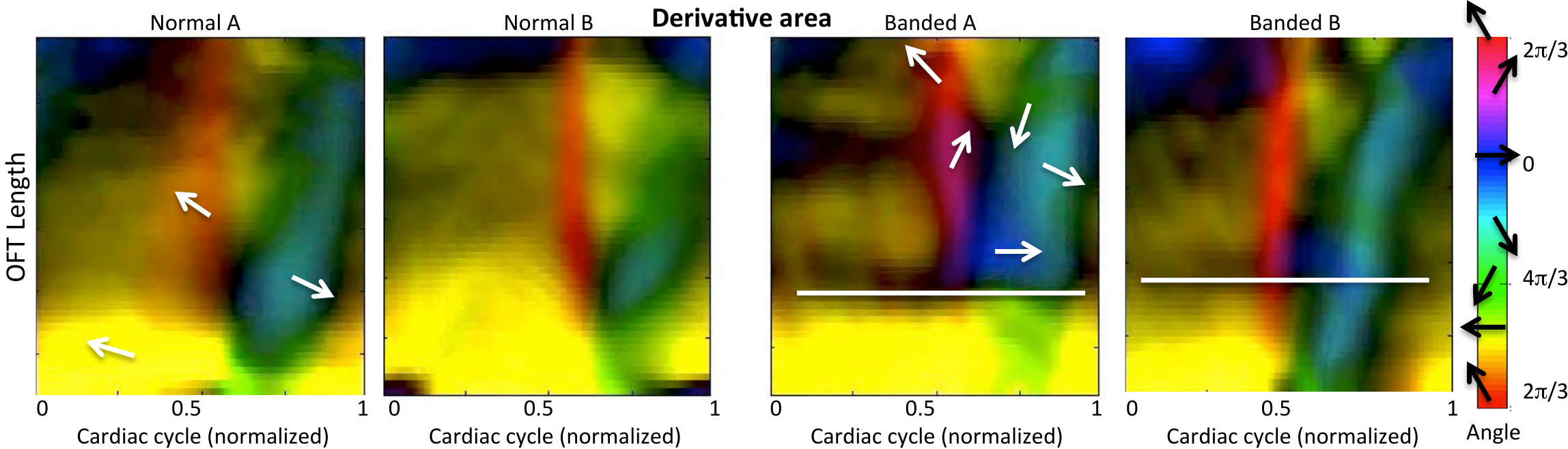}
\caption{Derivatives of the area plots. The direction of change is encoded in the color, the magnitude in the intensity (black to full color). Left is normal, right is banded. The areas where the expansion begins (red) and contraction ends (light blue) are clearly visible in the normal, but more disorganized in the banded. The banded image also clearly shows the band location (white line) where almost no contraction or expansion is happening.}
\vspace{-0.2in}
\label{fig:areaDeriv}
\end{center}
\end{figure}

From derivative plots of the 2D area images, the onset of the expansion and contraction regions are clearly visible in the images as diagonal lines (see Figure~\ref{fig:areaDeriv}).  
The most noticeable difference between the normal and the banded is that the banded are more disorganized, and have horizontal bands where the magnitude of the derivative is nearly zero.  

\subsection{OFT shape visualization and analysis}



\begin{figure}
\begin{center}
\includegraphics[width=\linewidth]{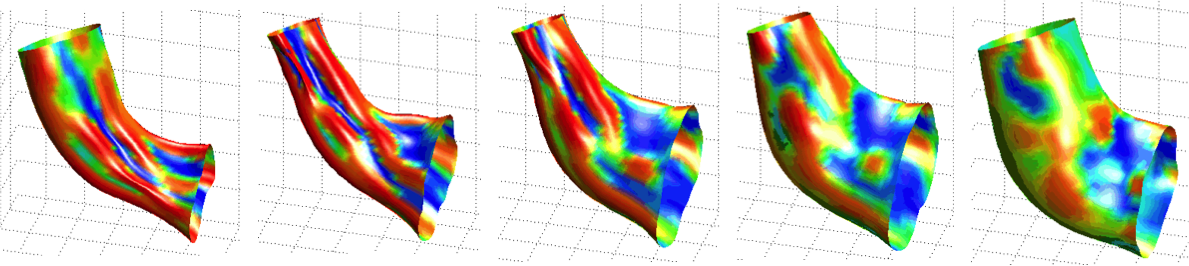}
\caption{Frames from a video showing the lumen at different states in the cardiac cycle, colored by mean curvature. See Supplemental for video.   }
\vspace{-0.2in}
\label{fig:lumenVideo}
\end{center}
\end{figure}

Plots of the lumen surface curvature over time allow visualization of the intricate motion of the endocardium over the cardiac cycle (see Figure~\ref{fig:lumenVideo}). Curvature allows easy visualization of folds that form on the endocardium as the OFT contracts, and how the endocardium unfolds upon OFT expansion. Curvature shows that many of the endocardial folds run through the entire length of the OFT. This is an interesting biological characteristic, as the folds are associated with tethering molecules that anchor the endocardium and myocardium together and promote folding and apposition of the endocardium upon OFT contraction.


The OFT myocardium is typically approximated as an elliptical shape that deforms to a circular one at maximum expansion. The myocardium circumferential curvature is more uniform at maximum expansion (see Figure~\ref{fig:OCT-OFT}b-e and~\ref{fig:crossVideo}). However, at maximum contraction and expansion it shows the same striations in roughly the same place. If the shape were truly elliptical there would be only two bands of high curvature, but instead there are multiple ones (from three to five). The normalized video also clearly shows these striations even at minimum contraction and maximum expansion (the images should go to all black and all white), indicating that the surface does not reach the curvature extremes at the same time. These bands also appear in the PCA analysis of the cross-sectional shape in the second mode of variation, where the arrows switch directions. The curvature values are consistent with a shape that has three to five ``corners'', with flatter areas between the corners. As the shape expands, the flatter areas bulge out faster, increasing their curvature, while the curvature at the corners actually decreases because it does not expand as fast as the adjacent tissue, flattening the corner out. 

There are no easily-quantifiable differences in curvature between the normal and the banded embryos. The third mode of variation in the PCA analysis appears to be more disorganized for the banded embryos (the arrow directions flip more times) and this is somewhat visible in the curvature videos as more disjoint regions with differing curvature values.




The strain images show the contraction wave onset and movement through the tube. In the banded example the band causes the shape to ``fold'' in the middle, inducing an additional radial band of strain. This band is persistent across the entire cycle (see supplemental images and Figure~\ref{fig:Strain}). 

\begin{figure}
\begin{center}
\includegraphics[width=\linewidth]{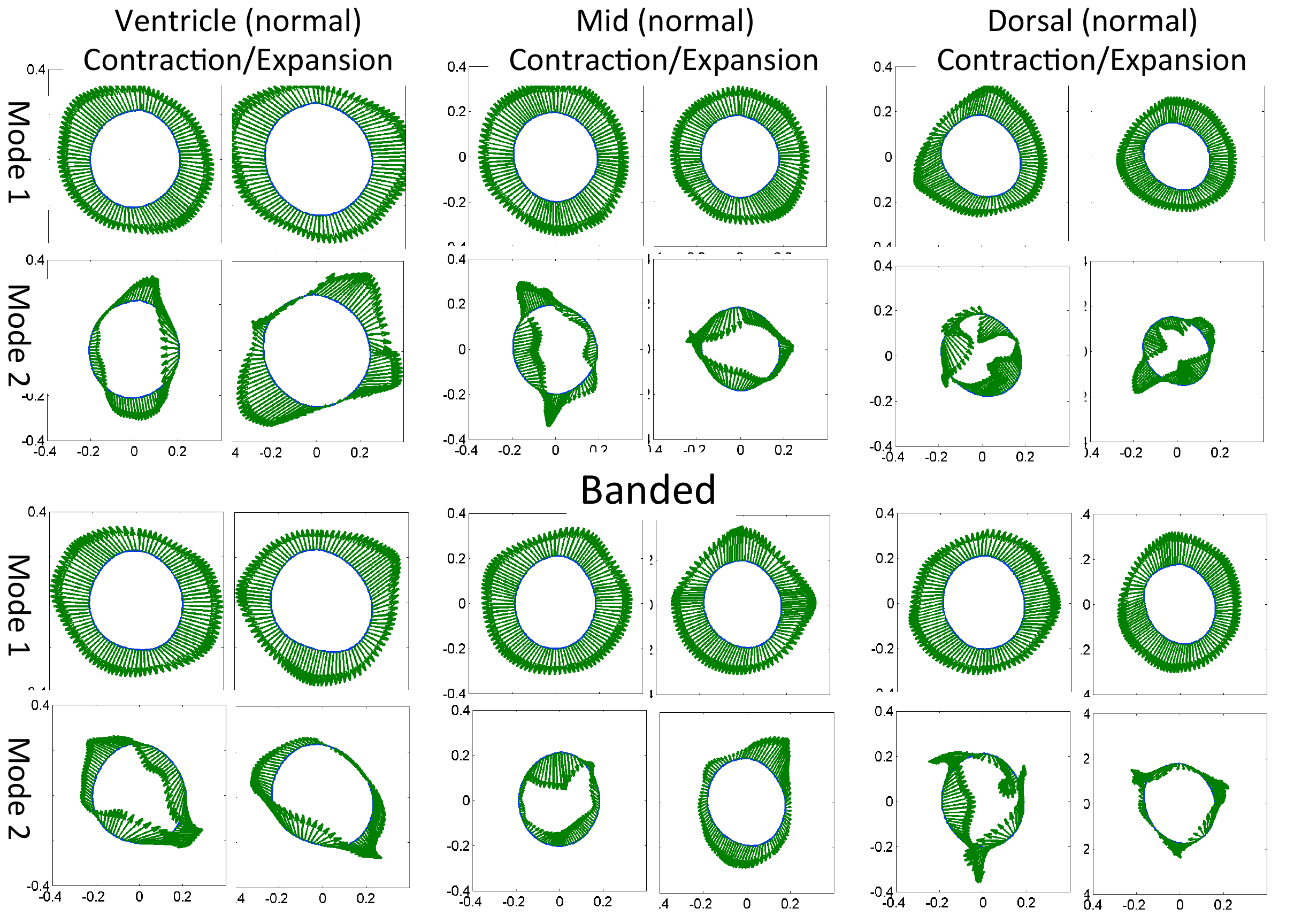}
\caption{PCA analysis of the shape change of three contours (ventricle, mid, dorsal), split into expansion and contraction phases of the cardiac cycle. Top: Normal, Bottom: Banded. Shown are the first two modes of variation.  The primary variation is an overall change in shape from expanded to contracted (and vice-versa). The second variation is orthogonal to the first, and essentially captures the difference in speed of contraction/expansion from the overall movement. }
\vspace{-0.2in}
\label{fig:SebastianPCA}
\end{center}
\end{figure}

By extracting cross-sectional contours and plotting them over time we can analyze the contraction/expansion motion of the OFT wall. Moreover, the PCA analysis enables a more clear depiction of expansion and contraction motions. In particular, our analysis show that expansion and contraction are not the same for all points on the contour, and that expansion is not symmetric with contraction. Figure~\ref{fig:SebastianPCA} shows the first and second modes of variation for a normal and a banded OFT, with the cardiac cycle split into the expansion and contraction phases. The visualization displays the movement direction for each point on the contour for that mode of variation. From this figure the effect of banding can be clearly seen: the expansion and contraction are more uniform in banded hearts and of much lower magnitude (roughly half that of the contours at either end) since the overall size is essentially constrained by the band. 

\section{Conclusion} \label{sec:discuss}

In this paper we have presented an algorithm for producing  consistent spatial correspondence for the deforming heart OFT and extending this correspondence (both temporally and spatially) to multiple embryos, within the limitations of the imaging data. This algorithm is suitable for aligning other tubal surfaces undergoing deformation. 

The primary purpose of the correspondence algorithm is to enable comparison within the cardiac cycle (single embryo) and between embryos. We have presented several visualization and analysis approaches that can be used to reduce the dimensionality of the data and quantify some aspects of the overall motion (e.g., peristaltic wave speed and cross sectional shape change). By mapping the shape, curvature, and area data to images we can employ computer vision techniques to further analyze cardiac motion at early stages of embryonic development.

\section*{Acknowledgments}

This research was funded in
part by NSF grants  DBI-1052688 and  IIS-1302142 and NIH grant  R01-HL094570. Any opinions,
findings, and conclusions or recommendations expressed in this
material are those of the authors and do not necessarily reflect the
views of the sponsors.

\bibliographystyle{elsarticle-harv}


\bibliography{VizHeart,chickHeart,prevWork-2-2.3,section2.4.2.4-crossParam,ferretBrain}

\end{document}